\documentstyle[prl,aps,twocolumn,floats]{revtex}
\begin{document}
\draft
\title{Brownian motion model of random matrices revisited}
\author{Sudhir R. Jain$^{\ddag }$ and Zafar Ahmed$^{\dagger}$   \\
$^{\ddag}$Theoretical Physics Division, Bhabha Atomic Research Centre,\\
Mumbai 400 085, India \\
$^{\dagger }$Nuclear Physics Division, Bhabha Atomic Research Centre,\\
Mumbai 400 085, India}
\maketitle

\begin{abstract}

We present a modified Brownian motion model for random matrices
where the eigenvalues (or levels) of a random matrix evolve in
``time" in such a way that they never cross each other's path.
Also, owing to the exact integrability of the level dynamics, we
incorporate long-time recurrences into the random walk problem
underlying the Brownian motion. From this model, we derive the
Coulomb interaction between the two eigenvalues. We further show
that the Coulomb gas analogy fails if the
confining potential, $V(E)$ is a transcendental function
such that there exist orthogonal polynomials with weighting function,
$\exp [-\beta V(E)]$, where $\beta $  is a symmetry parameter.
\end{abstract}


Nearly all physical systems that are of significance are quite
complicated due to their complex structure or interactions or a
combination of the two. Having complete knowledge about these
systems is unrealistic. As a result, statistical analyses where
we include as much salient information as is possible constitute
(pen-)ultimate theories. The well-known examples are some of the
most important branches of physics like statistical mechanics,
random matrix theory, information theory, and so on \cite{jain-alonso}.
Our interest here is in random matrix theory \cite{mehta} where a
complex or disordered quantum system is modelled in a realistic
statistical way by drawing its Hamiltonian from an ensemble of
random matrices. This mathematically elegant theory becomes physical
also as we incorporate dynamical features. The basic theme is to
develop a probabilistic (statistical) framework which is systematically
augmented by Newtonian or dynamical (deterministic) characteristics
that are fundamental attributes of the system. In this Letter, we
present a modified   Brownian motion model, originally proposed by
Dyson \cite{dyson62,dyson72}, that takes into account basic features
of level dynamics.

Matrix models are now employed to understand physical phenomena in varied
fields, particularly for  heavy nuclei \cite{brody}, mesoscopic disordered
conductors \cite{montambaux}, two-dimensional quantum gravity \cite{gross},
and chaotic quantum systems \cite{bohigas}.
A lot of excitement and activity in this field has been going on for nearly
two decades due to nearly-perfect modelling of universal
conductance fluctuations in metals \cite{washburn} and local
spectral fluctuations in chaotic quantum
systems \cite{bohigas}.

Given an $N \times N$ random matrix with eigenvalues, $E_1, E_2,...,E_N$,
the joint probabilty distribution function \cite{mehta}
\begin{mathletters}
\begin{eqnarray}
P(E_1,E_2,...,E_N) &=&
C_{N\beta }e^{ [-\beta W(E_1,E_2,...,E_N)]},  \\
W(E_1,...,E_N) &=& \sum_{j=1}^{N} V(E_j)-\sum_{i < j}
\log |E_i - E_j|
\end{eqnarray}
\end{mathletters}
gives the probability for eigenvalues to be lying  in unit intervals
around the points, $E_i$'s. This is also the probability density of
$N$ unit charges moving freely on the real line under the influence
of forces derived from the potential energy, (1b). Dyson \cite{dyson62,dyson72}
generalised the notion of matrix ensemble so that (1a) is the equilibrium
distribution of a statistical mechanical system. Here the many-body system
has fictitious particles as levels or eigenvalues and it evolves in
some fictitious time. In general, the dynamical system is in some
non-equilibrium state. This gives birth to the Brownian motion model
where eigenvalues, $E_i$ perform Brownian motion with no well-defined
velocities or inertia. The Coulomb gas is thus not a conservative system,
the particles are actually experiencing frictional forces that resist their
motion.

Two decades later, in an attempt to provide dynamical origin of (1a), it was
shown by Pechukas \cite{pechukas} and Yukawa \cite{yukawa} that the levels
can be thought to evolve with a parameter, $t$ (like ``time") if one considers
the Hamiltonian $H = H_0 + tV$ where $V$ is Hermitian. It was shown that
for eigenvalues, equations of motion can be written and the Hamiltonian
underlying level dynamics is, in fact, an exactly integrable generalised
Calogero-Moser system. There are two basic features that emerge out of these
recent works : (1) due to the fact that levels evolve adiabatically, there
should be no curves crossing each other as the system evolves and the
many-body system equilibrates; (2) since the system is an $N$-body Hamiltonian
system, in principle, Poincar\`{e} recurrences are bound to occur. Regarding
the recurrences, it is quite clear that as $N$ becomes large, the recurrence
time will be very large too. It is also quite obvious that {\it these two
important features are not satisfied in the Brownian motion model of Dyson}.
The aim of this Letter is to present a Brownian motion model where these
dynamical inputs are incorporated, in consonance with the opening discussion.

Consider $N$ levels or eigenvalues as random walkers which are initially
close together, walking on a line without crossing each other, with a
non-zero probability of recurrence after a long time. In this random walk problem,
we wish to find the probability distribution of positions of the random walkers
at some later time, $t$, given that there is recurrence at time, $2t$. At
time $t$, the position of an eigenvalue is $E_i(t), ~i=1,2,...,N$. Initially,
let the positions be $E_i(0)=E_{i,0}$ satisfying
\begin{equation}
-E_{\mbox{max}} < E_{1,0} < E_{2,0} < ... < E_{N,0} < E_{\mbox{max}}
\end{equation}
where $E_{\mbox{max}}$ is a fixed bound. The objective is to find the
joint probability distribution function for  levels
evolving without crossing each other's
path, i.e., maintaining the inequalities :
\begin{equation}
E_{i-1}(t')~ < ~E_i(t'), ~~~i=2,...,N, ~~\mbox{for}~~t' \leq t.
\end{equation}
Let us
denote the set of initial (final) set of levels, $(E_{1,0},...,E_{N,0})$
($(E_{1},...,E_{N})$) by $\tilde{E} _0$ ($\tilde{E}$).
Similar random walk problem is encountered in the discussion on commensurate
melting where domain walls move \cite{huse-fisher}. We use the arguments
presented in this work below to our purpose.
One can systematically treat this random walk problem on a lattice with
discrete time. Since we are interested in long times, we take recourse to
the continuum approximation where we have the simple case of Brownian motion.
The probability distribution of $N$ free, unrestricted random walkers is
\cite{itzykson-drouffe}
\begin{equation}
P_N^{0} (\tilde{E} _0,\tilde{E};t) =
\frac{e^{-\frac{\vert \tilde{E} - \tilde{E} _0 \vert ^2}{(2 Dt)}}}{(2\pi Dt)^{N/2}},
\end{equation}
where in consistency with the Brownian motion,
$D$ is the diffusion constant
that sets the time scale via
\begin{equation}
\langle (E_i - E_{i,0})^2 \rangle = Dt.
\end{equation}
The angular brackets denote the averaging with probability distribution.
The probability distribution satisfies the $N$-dimensional diffusion equation,
\begin{equation}
\sum_{j=1}^{N} \frac{\partial ^2P_N}{\partial x_j^2} = \frac{2}{D}
\frac{\partial P_N}{\partial t}.
\end{equation}
The distribution (4) must evolve in such a way that the linear
manifolds,
\begin{equation}
E_i = E_{i+1}, ~~i=1,2,...,N-1,
\end{equation}
are never crossed. Equivalently, the distribution $P_N = 0$ on
these manifolds. Since the manifolds (7) are linear, the problem can be
solved by the method of images. The levels are allowed to walk
unrestrictedly and randomly with positive and negative weights
starting at $\tilde{E}_0$ and its mirror images, according to (4).
If the initial weights are chosen such that $P_N$ is antisymmetric
under reflection in each manifold in (7), all subsequent distributions
will be antisymmetric too.

To obtain an analytic expression, we use the elements, $g$ of
$N$-dimensional  symmetric group, $S_N$ \cite{hamermesh}, to permute
the components of any $N$-tuple, $\tilde{E} = \{E_1,E_2,...,E_N\}$. To
obtain $P_N$, as remarked above, we start a level with weight,
$\epsilon (g)$ at each mirror-image point $g\tilde{E}_0$, where
$\epsilon (g) = +1$ ($-1$) for even (odd) permutations corresponding
to even (odd) number of reflections.
For $\tilde{E}$ satisfying (3), we have
\begin{mathletters}
\begin{eqnarray}
P_N (\tilde{E} _0,\tilde{E};t) &=&
\sum_{g \in S_N} \epsilon (g)P_N^{0} (g\tilde{E} _0,\tilde{E};t) \\
&=& {\cal A}_N
\frac{e^{\frac{-\vert \tilde{E} \vert ^2 - \vert \tilde{E} _0
\vert ^2}{(2 Dt)}}}{(2\pi Dt)^{N/2}},
\end{eqnarray}
\end{mathletters}
where ${\cal A}_N$ is the antisymmetrised sum,
\begin{equation}
{\cal A}_N(\tilde{E} _0, \tilde{E}; t)
= \sum_{g \in S_N} \epsilon (g) e^{\frac{\tilde{E} .g\tilde{E} _0}{Dt}}.
\end{equation}
Let us now define the recurrence of fictitious particles or the reunion of
the walkers around $\overline{E}$ by
\begin{equation}
\overline{E} - E_{\mbox{max}} < E_1 < E_2 < ... < E_N < \overline{E} + E_{\mbox{max}}
\end{equation}
where we assume (without loss of generality) that the initial and final mean
positions are the origin and $\overline{E}$. Thus,
$\tilde{E}.g\tilde{E} _0 = \sum_{j=1}^{N} (E_{j} - \overline{E})(g\tilde{E} _0)_j
={\cal O}(E_{\mbox{max}}^2)$.
Due to antisymmetrisation, the form taken by ${\cal A}_N$ contains the
Vandermonde determinants in initial and final components. That is,
\begin{eqnarray}
{\cal A}_N(\tilde{E} _0, \tilde{E}; t) &=&
\frac{(Dt)^{-n_N}}{1!2!...(N-1)!} \prod_{j<k}^{N} \vert E_{j,0}-E_{k,0}\vert
\vert E_{j}-E_{k}\vert \nonumber \\
&+& {\cal O}\left(\frac{E_{\mbox{max}}^2}{Dt} \right),
\end{eqnarray}
$n_N = {1 \over 2}N(N-1)$. The asymptotic probability for recurrence is,
in fact, from (8b) and (11), given by
\begin{eqnarray}
P_N(\tilde{E} _0,\tilde{E};t) &=& \frac{e^{-\frac{N}{2Dt}\overline{E}^2}}{(2\pi )^{N/2}(Dt)^{N^2/2}}
\prod_{j<k}^{N} \vert E_{j,0}-E_{k,0}\vert \vert E_{j}-E_{k}\vert \nonumber \\
&~&\frac{1+{\cal O}\left(\frac{E_{\mbox{max}}^2}{Dt} \right)}{1!2!...(N-1)!}.
\end{eqnarray}
Writing (12) as the partition function for $N$ fictitious particles, we
obtain the Coulomb interaction. To emphasise this important result, we have
shown how {\it the Coulomb gas analogy follows  directly from a  conservative
(neighbors being always the same), recurring Brownian motion model}.

Any two eigenvalues interact with each other by the Coulomb interaction
in two dimensions. The potential that confines these eigenvalues, $V(E)$
is the one we require now to complete our discussion. The universality of
local spectral fluctuations is due to the fact that the Coulomb gas analogy
holds true. The non-universal aspect is the average level density,
$\sigma (E)$, which is connected with the confining potential by the
following equation \cite{dyson72}:
\begin{equation}
{\cal P}\int_{-\infty}^{\infty } dE ^\prime \frac{\sigma (E ^\prime )}{E - E ^\prime} =
\frac{\partial V(E)}
{\partial E}.
\end{equation}
The left hand side is a principal value integral.

With the confining potential thus found, one can explore the situations
where for the weighting function, $\exp [-\beta V(E)]$, there exist
orthogonal polynomials. This enables one to use the powerful method
of orthogonal polynomials in finding various correlations between the
eigenvalues \cite{mehta}.
We now come to our second result in the form of a theorem.

\noindent
{\bf \sf Theorem :~} If the matrix elements of a random matrix are
distributed by a transcendental function so that the confining potential
of eigenvalues in the continuum approximation is also a transcendental
function $V(E)$, then the  Coulomb gas model breaks down if
there  exist orthogonal polynomials with a weighting function
$\exp [-\beta V(E)]$.

\noindent
{\bf \sf Proof :~}
Using {\it Reductio ad absurdum}, let us  assume  that the
Coulomb gas  model works.
Let the points, $E_1,E_2,E_3,...$ be such as to minimize the potential energy,
\begin{equation}
W=\sum_{j=1}^{N} V(E_j) - \sum_{i\le j } \log |E_i-E_j|.
\end{equation}
Then,
\begin{equation}
0=-{\partial W \over \partial E}(E_j)=- \frac{\partial V}{\partial E}(E_j)
+ \sum_{i\ne j} {1 \over E_j-E_i}.
\end{equation}
Let us assume that there exist orthogonal polynomials,
$p_N(E)$, with $e^{-\beta V(E)}$ as
the  weighting function.
Let $E_j$ be the zeros
of $p_N(E)$. By the fundamental theorem of algebra,
\begin{equation}
p_N(E)=C (E-E_1)(E-E_2)(E-E_3).....(E-E_N),
\end{equation}
$C$ being a  constant.
Next, we can write \cite{mehta}
\begin{equation}
{p^{\prime \prime}_N(E_j) \over 2 p^{\prime}_N(E_j)}=\sum_{i\ne j} {1 \over E_i-E_j}.
\end{equation}
Substituting in (15), we get a differential equation for $p_N(E)$,
\begin{equation}
{p^{\prime \prime}_N(E_j) \over 2 p^{\prime}_N(E_j)}-
\frac{\partial V}
{\partial E} (E_j)=0,
\end{equation}
which is absurd for a ratio of two polynomials can never be a
transcendental function.
This concludes the proof.

The Brownian motion model has served as a basis for most of the
developments in quantum transport theory. In this Letter, we have pointed
out the features that are now well-known from the results in level
dynamics. Since  Dyson's Brownian motion model does not satisfy these
requirements, we have given a modified model. This has the most
important distinction of giving us the Coulomb gas model as a result
of the constraint that the levels cannot cross each other. Thus the
Fermionic nature of levels emerges from the random walk problem
considered.

Moreover, we have proved that the Coulomb gas analogy breaks down
for the case when the confining potential is a transcendental function
such that there exist orthogonal polynomials with the corresponding
weighting function. However, even if the Coulomb gas model is inconsistent
with any transcendental confinement, if the joint probability distribution
function (j.p.d.f.) satisfies the $N$-dimensional diffusion equation, the local
correlations may still be universal.
This situation arises in mesoscopic disordered conductors
\cite{ahmed-jain} where the
conductance fluctuations are universal and the confining potential is
$\log ^2(E)$. The Coulomb gas analogy breaks down there, however the
j.p.d.f. satisfies the diffusion equation.

\end{document}